\documentclass[12pt]{article}
\usepackage{amsfonts,amsthm,amsmath,amssymb}
\usepackage{hyperref}
\usepackage{cite}
\usepackage[paper=letterpaper,margin=1in]{geometry}
\usepackage{graphicx}
\usepackage{units}

\def\a'{\alpha'}

\renewcommand\bar{\overline}

\newcommand{\be}{\begin{equation}}
\newcommand{\ee}{\end{equation}}
\newcommand{\bea}{\begin{eqnarray}}
\newcommand{\eea}{\end{eqnarray}}
\newcommand{\ba}{\begin{array}}
\newcommand{\ea}{\end{array}}
\newcommand{\bee}{\begin{enumerate}}
\newcommand{\eee}{\end{enumerate}}
\newcommand{\bi}{\begin{itemize}}
\newcommand{\ei}{\end{itemize}}
\newcommand{\bc}{\begin{center}}
\newcommand{\ec}{\end{center}}
\newcommand{\bfig}{\begin{figure}}
\newcommand{\efig}{\end{figure}}

\numberwithin{equation}{section}

\begin{document}
\thispagestyle{empty}
\renewcommand{\thefootnote}{\fnsymbol{footnote}}

\begin{flushright}
\end{flushright}

\begin{center}

\bf{{\Large Holographic Entanglement Entropy for\\

\medskip

Gravitational Anomaly in Four Dimensions}

\vspace{1cm}}

{Tibra Ali$^{a}$\,\footnote[4]{\tt{tibra.ali@pitp.ca}}, S. Shajidul Haque$^{b}$\,\footnote[3]{{\tt{sheikh.haque@uct.ac.za}}}, Jeff Murugan$^{b,c}$\,\footnote[1]{\tt{jeff.murugan@uct.ac.za}}}\\

\vspace{0.5 cm}\vspace{0.2 cm}{{\it$^{a}$Perimeter Institute for Theoretical Physics \\31 Caroline Street N., Waterloo, ON N2L 2Y5, Canada} }

\vspace{0.5 cm}{{\it$^{b}$Laboratory for Quantum Gravity \& Strings\\
Department of Mathematics \& Applied Mathematics\\
University of Cape Town,
South Africa }} \\

\vspace{0.5 cm}{{\it$^{c}$School of Natural Sciences\\
Institute for Advanced Study\\
Princeton, NJ 08540, USA}} \\

\vspace{2cm}
{\bf Abstract}
\end{center}
\begin{quotation}
\noindent
We compute the holographic entanglement entropy for the anomaly polynomial $\mathrm{Tr} R^2$ in 3+1 dimensions. Using the perturbative method developed for computing entanglement entropy for quantum field theories, we also compute the parity odd contribution to the entanglement entropy of the dual field theory that comes from a background gravitational Chern-Simons term. We find that, in leading order in the perturbation of the background geometry, the two contributions match except for a logarithmic divergent term on the field theory side. We interpret this extra contribution as encoding our ignorance of the source which creates the perturbation of the geometry.
\end{quotation}

\setcounter{page}{0}
\setcounter{tocdepth}{2}
\newpage
 \tableofcontents
\section{Introduction}
The idea that space and time may be emergent macroscopic properties obtained by coarse-graining over some microscopic quantum substructure is by no means a new one \cite{Seiberg:2006wf, Koch:2009gq}. It is an idea with a rich history and an equally interesting geography in the landscape of theoretical physics. Perhaps the most concrete laboratory in which this view of nature may be tested is furnished by the gauge/gravity duality and, more specifically, Maldacena's AdS/CFT correspondence \cite{Maldacena:1997re}. In this context, the geometrical properties of a dual classical spacetime are seen to emerge from the large-$N$ dynamics of a quantum field theory of interacting $N\times N$ matrices. But even in this setting, understanding the emergent properties of realistic, cosmological or black hole spacetimes has remained out of reach. This is due, in no small part, to the fact that such spacetimes necessitate the relaxation of the rigid constraints of supersymmetry and integrability. Without the control afforded by non-renormalization theorems, many of the computations cannot be trusted to give sensible answers. Questions abound: Is there a concrete mechanism for emergence? How do we build realistic spacetimes? What are the relevant quantities to study even?

Fortunately, hope is on the horizon. Literally. The past decade has seen several promising developments in our understanding of  black holes  and the information loss problem they pose. Much of this understanding is intimately related to the idea of {\it entanglement} in the quantum field theory dual to the black hole geometry. In a sense, quantum entanglement has emerged as a glue of sorts, with which to put together quantum states and eventually produce a classical spacetime. Entanglement, as characterized by its various associated entropies is, of course, a notoriously difficult quantity to compute in quantum field theories in general and gauge theories in particular. This situation was changed in a fairly dramatic way recently by the celebrated work of Ryu and Takayanagi which relates the entanglement entropy of some boundary field theory to the area of an open minimal surface in the bulk whose boundary is set by the entangling surface \cite{Ryu}. Since its discovery, the Ryu-Takayanagi formula has provided one of the most useful tools in contemporary gauge/gravity problems. And, like any tool, figuring out where and how it fails is just as important as understanding why it works.

Recently, building on the work of Lewkowycz and Maldacena \cite{Lewkowycz}, Dong has proposed a generalization of the Ryu-Takayanagi prescription for computing the holographic entanglement entropy in a general field theory dual to a higher derivative gravity theory whose Lagrangian is constructed from contractions of the Riemann tensor \cite{Dong}. Accordingly, in $d$-dimensions Dong's proposal reads\footnote{The notation in equation (\ref{eq:Dong-covariant}) is explained in more details in Dong's paper \cite{Dong}. Later in our paper, we explain and use a non-covariant version of Dong's formula (\ref{eq:Dong-covariant}) which is also taken from \cite{Dong}.}
\begin{eqnarray} \label{eq:Dong-covariant}
  S_{\mathrm{EE}} &=& 2\pi \int\,d^{d}y\,\sqrt{g}\,\Biggl\{ -\frac{\partial L}{\partial R_{\mu\nu\rho\sigma}\varepsilon_{\mu\nu}
  \varepsilon_{\rho\sigma}} + 2\sum_{\alpha}\left(\frac{\partial^{2}L}{\partial R_{\mu_{1}\nu_{1}\rho_{1}\sigma_{1}}
  \partial R_{\mu_{2}\nu_{2}\rho_{2}\sigma_{2}}}\right)_{\alpha}\frac{K_{\lambda_{1}\nu_{1}\sigma_{1}}
  K_{\lambda_{2}\nu_{2}\sigma_{2}}}{q_{\alpha}+1}\times\nonumber\\
  &\times& \Bigl[\left(n_{\mu_{1}\mu_{2}}n_{\rho_{1}\rho_{2}} - 
  \epsilon_{\mu_{1}\mu_{2}}\epsilon_{\rho_{1}\rho_{2}}\right)n^{\lambda_{1}\lambda_{2}} + \left(n_{\mu_{1}\mu_{2}}n_{\rho_{1}\rho_{2}}
   + \epsilon_{\mu_{1}\mu_{2}}\epsilon_{\rho_{1}\rho_{2}}\right)\epsilon^{\lambda_{1}\lambda_{2}}
  \Bigr]\Biggr\},
\end{eqnarray}
and corrects Wald's gravitational entropy formula (the first term above), by accounting for extrinsic curvature terms crucial to its interpretation as entanglement 
entropy. These terms, in turn, can be interpreted as `anomalies' in the variation of the action. Where they coincide these two formulations agree, since the extrinsic curvature vanishes on the Killing horizon. Dong's proposal, however, goes much further and properly reproduces the entropy in a variety of higher derivative theories of gravity including, $f(R)$ gravity, general four-derivative gravity, Lovelock gravity and, in 3-dimensions, topologically massive gravity \cite{Castro}. Other significant attempts at generalizing the holographic entanglement entropy proposal to higher-derivative gravity include \cite{deBoer,Hung,Camps,Myers,Bhattacharyya,Mozaffar}

In this article, we test the proposal against another novel, higher derivative theory, namely a Chern-Simons modification of 4-dimensional Einstein gravity \cite{Jackiw:2003pm}. Originally proposed by Jackiw and Pi as a phenomenological extention of general relativity obtained by lifting the 3-dimensional gravitation Chern-Simons term
\be
  \Omega_{3}(\Gamma) = \mathrm{Tr}\left(\Gamma\wedge d\Gamma + \frac{2}{3}\Gamma\wedge\Gamma
  \wedge\Gamma\right),
\ee
to four spacetime dimensions, the theory breaks CPT as well as diffeomorphism invariance. While the former is manifest in the action, the latter is hidden. Consequently, 4-dimensional Chern-Simons modified gravity still propagates two physical degrees of freedom and gravitational waves still possess two polarizations. Many, although not all\footnote{Among those that are not solutions of the Chern-Simons modification are the Kerr, Kerr-Newman and Kerr-NUT spacetimes.}, of the solutions of general relativity, including the Schwarzschild and pp-wave metrics, persist in the Chern-Simons deformation. However, in this article we assume that the coefficient of the Chern-Simons term in the four-dimensional action is a constant and in that case this term becomes topological. In the context of Riemannian geometry this terms is known as the Chern-Pontraygin density. In the physics context, however, this is an anomaly polynomial which has many applications starting from gravitational instantons to anomalies in the gauge or the gravity sector \cite{Alvarez-Gaume}. In the spirit of \cite{Alvarez-Gaume} we call this polynomial a `gravitational anomaly' in the title of this paper although there are no pure gravitational anomaly in four-dimensions. This article concerns itself with the question: {\it What effect does the addition of a Chern-Pontryagin density to the Einstein-Hilbert action have on the entanglement entropy?} There is, of course, excellent reason to suppose that it does in fact have an effect. The Chern-Pontryagin density is, after all, topological in nature and so we would certainly expect it to contribute to the universal terms of the entanglement entropy \cite{Kitaev:2005dm}. How exactly, is the subject of this note.

\section{Entropy Functional for the Anomaly Polynomial}
The bulk theory that we are interested in is given by the following action
\be
I= -\frac{1}{16\pi G}\int d^4 x \sqrt{g}\left(R+\frac{2}{\ell^2}-\frac{\kappa}{4} *\!RR\right) , \label{eq:action1}
\ee
where
$*RR\equiv *(R^M{}_{N}){}^{PQ} R^N{}_{M} {}_{PQ} \equiv\frac{1}{2} \ \epsilon^{PQST} R^M {}_{NST} R^M{}_{NPQ} $ is the Chern-Pontryagin density. The cosmological constant is $\Lambda = -\frac{1}{\ell^2}$, where $\ell$ is the length-scale corresponding to the AdS space. This action (\ref{eq:action1}) is a special case of the so-called Chern-Simons modified gravity theory introduced in \cite{Jackiw:2003pm}. In that paper, however, the coefficient of the $*RR$ term is a spacetime dependent scalar. It makes a contribution to the bulk equations of motion which involves the four-dimensional Cotton tensor. In our case, we take $\kappa$ to be a constant and the anomaly form, which is topological in nature, makes no contribution to the bulk equations of motion. As a result, four dimensional anti-de Sitter space continues to be the maximally symmetric solution of this modified theory. The CP violation effects of this term was explored in \cite{Deser5}.

In the absence of the anomaly term, the contribution to the holographic entanglement entropy contribution is given by the Ryu-Takayanagi formula first proposed in \cite{Ryu}. This proposal was later proved in \cite{Lewkowycz} using the holographic replica trick. Recently, Dong \cite{Dong} has extended the results of \cite{Lewkowycz} for theories that contain higher derivative corrections as in (\ref{eq:action1}).

The essence of \cite{Lewkowycz,Dong} is to use the holographic replica trick in the following way. Let $M_1$ be some spacetime manifold with some conformal field theory living on it. We shall often refer to $M_1$ as the boundary manifold as we shall think of it as being the boundary of some bulk geometry. We assume that the CFT is in its ground state. We then want to compute the entanglement entropy of the ground state in some region $A$ of $M_1$ with its complement $\bar{A}$. To do this, we then extend $M_1$ (assumed to be static and analytically continued to Eulcidean signature) to its $n$-fold cover $M_n$ by excising the region $A$ along its boundary $\partial A$ and glueing $n$ copies of $M_1$ along $\partial A$ in a cyclic fashion. The group $\mathbb{Z}_n$ permutes each component of $M_n$. The R\'enyi entropy of this system is given by
\begin{align}
S_n &= -\frac{1}{n-1}\log{\mathrm{Tr}[\rho^n]} \, ,
\end{align}
where $\rho$ is the density matrix associated with the region $A$ of the original manifold $M_1$. The R\'enyi entropy can be rewritten as
\begin{align}
S_n=-\frac{1}{n-1} \left(\log{Z_n} - n \log{Z_1}\right)\, , \label{eq:Renyi}
\end{align}
where $n>1$, $Z_n$ and $Z_1$ are the partition functions of the CFTs on $M_n$ and $M_1$, respectively. We are interested in the von Neumann entropy of the CFT on $M_1$. Formally, this is computed by the analytical continuation of (\ref{eq:Renyi}) to $n\rightarrow 1$. 

It is worth noting, however, that $M_n$ doesn't have a geometric interpretation for non-integer values of $n$ and so it is not clear what the above analytic continuation means geometrically for the boundary manifold. It was pointed out in \cite{Lewkowycz} that for theories that have holographic duals, the bulk geometry $B_n$ (which is the bulk geometry associated to the replica manifold $M_n$) does indeed have a geometric meaning even for non-integer values of $n$. $B_n$ is completely regular but the action of $\mathbb{Z}_n$ on it has a fixed point set $C_n$ which is a codimension-2 surface. The orbifold bulk geometry $\hat{B}_n = B_n/\mathbb{Z}_n$, whose boundary is our original boundary manifold $M_1$, thus has a singular surface $C_n$ with a conical deficit of $\epsilon = 1 - \nicefrac{1}{n}$. One can regularize this cone by introducing a smoothing parameter $a$, and the metric of the manifold near this surface is given by \cite{Loganayagam}
\begin{align}
ds^2 &= e^{2 A} \left[dz d\bar z +e^{2 A} T (\bar z dz-z d\bar z)^2\right] +(g_{ij} + 2 K_{aij} x^a+Q_{abij} x^a x^b ) dy^i dy^j \nonumber \\
&+ 2 i e^{2 A} (U_i +V_{ai} x^a ) (\bar z dz -z d\bar z)dy^i+\dots \, , \label{eq:cone}
\end{align}
where $x^a= \{z, \bar z\}$ are the complexified coordinates transverse to the codimension-2 surface and $y^i, \mbox{ with }i=1,2,... d$, are the coordinates along this surface. The functions, $T,g_{ij}$, $K_{aij}$, $Q_{abij}$, $U_i$, $V_{ai}$ all depend on $y^i$. $A= -\frac{\epsilon}{2} \ \text{log} (|z|^2+a^2)$ is the regularization function  that smooths out the squashed cone. The regularization parameter $a$ keeps track of the contribution from the singular limit of the cone (when $a\rightarrow 0$), which is subtracted  before taking the $a\rightarrow 0$ limit.

The holographic formula (in the large $N$ limit) for the entanglement entropy then is given by
\begin{align}
S_{\mathrm{EE}}=\lim_{n\rightarrow 1} \frac{n}{n-1}\left(S[\hat{B}_n] - S[\hat{B}_1]\right)=\left.\partial_n S[\hat{B}_n]\right|_{n=1} \, ,
\end{align}
where $S[\hat{B}_n]$ and $S[\hat{B}_1]$ are the classical bulk action evaluated on the orbifolds $\hat{B}_n$ and $\hat{B}_1$, respectively. As noted above, since the fixed point set $C_n$ is singular on the orbifolds, we smooth out the geometry near the tip of the cone by excising a small region and replacing the tip of the cone by a smoothed-out tip. Calling this new smoothed-out region near the tip as the `inside' region, it can then be shown that \cite{Dong}
\begin{align}
S_{\mathrm{EE}} = -\left. \partial_\epsilon S[\hat{B}_n]_{\mathrm{inside}}\right|_{\epsilon=0}.
\end{align}
Applying this to theories whose bulk gravitational action contains only the Einstein-Hilbert term (in addition to the usual cosmological constant) yields the usual Wald term in the expression for the entanglement entropy. It was shown by Dong \cite{Dong} that for theories with higher derivative coordinate-invariant terms one gets a correction to the Wald term. The total entanglement entropy then is given by (\ref{eq:Dong-covariant}) when expressed in a coordinate-invariant way. For explicit computations, however, it is convenient to express (\ref{eq:Dong-covariant}) in the coordinate system implicit in (\ref{eq:cone}). One then gets the following expression for the holographic entanglement entropy
\begin{align}
S_{\mathrm{EE}}=2\pi \int_{\Sigma} d^dy \sqrt{g}\left\{\frac{\partial L}{\partial R_{z\bar{z}z\bar{z}}}+\sum_\alpha\left(\frac{\partial^2 L}{\partial R_{zizj} \partial R_{\bar{z}k\bar{z}l}} \right)\frac{8K_{zij} K_{\bar{z}kl}}{q_\alpha+1}\right\} \label{eq:Dong}\, ,
\end{align}
where the integral is taken over a codimension-2 surface $\Sigma$ that is homologous to the entangling surface in the dual CFT on the boundary. The extra terms derived by Dong arise from would-be logarithmic divergences that come from the squashed cone method. Consequently, these are naturally interpreted as anomaly terms and the coefficients $q_\alpha$ can be thought of as `anomaly coefficients.' In our case, however, the anomaly coefficient is trivial. We refer the reader to \cite{Dong} for a more detailed discussion on this issue.

A new issue that arises with the addition of the new terms is how to determine the entangling surface in the bulk. The rigorous way of deriving the entangling surface is to solve the equations of motion. But this could be too difficult in practice and so Dong \cite{Dong} conjectures that the appropriate surface is the one that extremizes (\ref{eq:Dong}). Since in our case, the anomaly polynomial does not add any new term to the equations of motion, the bulk entangling surface will be same as the Ryu-Takayanagi surface. 

In order to understand how Dong's prescription gives the expression for holographic entanglement entropy, let us denote the two relevant part of the Lagrangian (\ref{eq:action1}) by
\be
L_1 = - \frac{1}{16\pi G} R \, , \ L_2  = \frac{\kappa}{64\pi G} *RR\, .
\ee
According to \cite{Dong}, the contribution that $L_1$ makes to the entanglement entropy is given by
\be
\lim_{\epsilon\rightarrow 0} \frac{1}{4G} \int d^4x \sqrt{G}\left(\frac{\delta^2(x^1,x^2)}{(\rho^2 + a^2)^\epsilon} - \frac{\epsilon \log(\rho^2+a^2)}{(\rho^2 +a^2)^\epsilon} \delta^2(x^1,x^2)\right) \, ,
\ee
where $\rho$ is the polar coordinate defined by $\rho=|z|$. In the $\epsilon \rightarrow 0 $ limit the second term drops out and it is easy to see that the first term is nothing but the Ryu-Takayanagi formula
\be
\left. S^{(1)}_{\mathrm{EE}}\right|_{L_1}= \frac{\mathcal{A}}{4G}\, ,
\ee
where
$\mathcal{A}= \int_\Sigma d^2y \sqrt{g}.$
Since the first term $L_1$ is first order in curvature it doesn't make any contribution to the entanglement entropy coming from the second term in (\ref{eq:Dong}). $L_2$, on the other hand, contributes to both. Its contribution to the Wald term is computed to be
\begin{align}
\left. S^{(1)}_{\mathrm{EE}}\right|_{L_2}= - \frac{\kappa}{4G}\int d^2y \sqrt{g}\left(\frac{\partial U_j}{\partial y_i}-\frac{\partial U_i}{\partial y_j}+2 g^{kl} K_{z\; jk} K_{\bar{z}\;il} \right) \epsilon^{ij}.
\end{align}
The contribution from the $L_2$ term to the second term in (\ref{eq:Dong}) can be shown to be
\be
\left. S^{(2)}_{\mathrm{EE}}\right|_{L_2}= -\frac{\kappa}{2G}\int d^2y\sqrt{g} K_{z\; ij} K_{\bar{z}\; kl} \epsilon^{jl} g^{ki},
\ee
which cancels out the second term in $L_2$'s contribution to $S^{(1)}_{\mathrm{EE}}$. Thus the final result is (continuing back to Lorentzian signature)
\be
S_{\mathrm{EE}} = \frac{1}{4G_N}\int d^2 y \sqrt{g} - \frac{\kappa}{4G}\int d^2 y \sqrt{g}\left[\partial_i U_j - \partial_j U_i\right] \epsilon^{ij}. \label{eq:ee-coordinates}
\ee
This expression is computed in the particular coordinate system given above. This can be expressed in the following coordinate-invariant way
\be
S_{\mathrm{EE}} = \frac{1}{4G} \int_\Sigma d^2y \sqrt{g} + \frac{\kappa}{8G}\int_\Sigma F_{ij} dy^i \wedge dy^j \label{eq:entropyformula} \, ,
\ee
where $F_{ij} = \partial_i A_j -  \partial_j A_i$ is the curvature of the normal bundle of $\Sigma$. The Abelian connection $A_i$ of the normal bundle is given by
\be
A_i = - \frac{1}{2} \epsilon_a{}^b\, \Gamma^a_{i b}. \label{eq:gauge}
\ee
The early Latin indices denote the normal direction to the surface $\Sigma$ and in four spacetime dimensions they take on two values. The quantity $\Gamma^a_{ib}$ is given by
\begin{align}
\Gamma^a_{ib} = \left(\partial_i n^\mu_b+ \hat{\Gamma}^\mu_{\sigma \nu} e^\sigma_i n^\nu_b\right) n^a_\mu \, ,
\end{align}
where $\hat{\Gamma}^\mu_{\sigma\nu}$ are the Christoffel symbols of the bulk spacetime and $e^\sigma_i =\frac{\partial x^\sigma}{\partial y^i}$ is the pull-back map. $n^a_\mu$ for $a=0, 1$ are the unit normal vectors. In our convention $n^0$ is time-like, while $n^1$ is space-like. Happily, this computation reproduces the results given in \cite{Loganayagam}.

\section{Holographic Entanglement Entropy}

In the previous section we have seen that the entanglement entropy of a bulk theory that contain an additional Chern-Pontryagin term is given by the usual Ryu-Takayanagi term and an additional term in (\ref{eq:entropyformula}) that computes the flux of the curvature of the normal bundle through the bulk entangling surface. As argued in the previous section, adding new terms in the action can in principle change the criterion for the bulk entangling surface. But since in our case the extra term is topological, we see that the bulk entangling surface coincides with the one prescribed by Ryu and Takayanagi.

It then follows that the contribution from the first term in (\ref{eq:entropyformula}) would be identical to the case that one would get if the bulk theory had just the Einstein-Hilbert and the cosmological constant terms. The new contribution would then come from the second term:
\be
\Delta S_{\mathrm{EE}}= \frac{\kappa}{8G}\int_\Sigma F_{ij}\, dy^i \wedge dy^j \label{eq:entropycorrection}\, ,
\ee
where $\Sigma$ is the Ryu-Takayanagi surface. Even though this expression was derived using Dong's prescription \cite{Dong}, it could have also been derived on dimensional grounds based on the fact that it computes the flux of the normal bundle through a codimension-2 surface in four spacetime dimensions. See \cite{alvaro} for more details.

This new term, (\ref{eq:entropycorrection}), is itself topological in the sense that the integrand is an exact form which, by Stokes' theorem, can be written as
\begin{align}
\Delta S_{\mathrm{EE}} = \frac{\kappa}{8G} \int_{\partial \Sigma} A_i dy^i \, .
\end{align}
One can now express this formula in terms of the bulk coordinates $x^\mu$ by introducing a new bulk gauge field $a_\mu := A_i  \frac{\partial y^i}{\partial x^\mu}$. In terms of this gauge field the above term becomes
\begin{align}
\Delta S_{\mathrm{EE}} = \frac{\kappa}{8G} \int_{\partial \Sigma} a_\mu dx^\mu\, .
\end{align}
From the definition of $A_i$ in (\ref{eq:gauge}) one can easily show that $a_\mu$ is given by
\begin{align}
a_\mu = -\frac{1}{2}\epsilon_a{}^b (\nabla_\mu n^\rho_b) n^a_\rho \, , 
\end{align}
where $\nabla$ is the Levi-Civita connection of the bulk metric. 

The boundary contour $\partial \Sigma \subset \partial B_1=M_1$ and so we can express $\Delta S_{\text{EE}}$ in terms of quantities that are intrinsic to $M_1$. We shall assume that the projections of the normal vectors $n^\mu_a$ on the boundary coincides with the values of those vectors at the boundary. Thus, if $m_\mu$ is the space-like normal to the boundary $\partial\Sigma$ and $h_{\mu\nu}= G_{\mu\nu} - m_\mu m_\nu$ is the induced metric on the boundary, then
\be
\left. h^\mu_\nu\, n^\nu_a\right|_{\partial\Sigma} = \left. n^\mu_a\right|_{\partial\Sigma}\, .
\ee
In other words, $n^a_\mu m^\mu =0$. We define hatted quantities to be the projection of a bulk quantity onto the boundary, $\hat{X}^\mu=h^\mu_\nu X^\nu$. With the above assumption we can write down the boundary projection of the gauge field as follows:
\be 
\hat{a} 
=-\frac{1}{2}\epsilon_a{}^b (\hat{\nabla}_\nu \hat{n}^\rho_b) \hat{n}^a_\rho\, . \label{eq:boundary-gauge}
\ee
We want to emphasize that this explicit expression for $\hat{a}$ only holds if the normal vectors at the boundary are orthonormal to $m^\mu$. Finally, the contribution to the entanglement entropy for the gravitational Chern-Pontryagin term expressed in terms of the projected gauge field $\hat{a}_\mu$ is simply
\be
\Delta S_{\text{EE}} = \frac{\kappa}{8G} \int_{\partial \Sigma} \hat{a}_\mu \hat{dx}^\mu\, . \label{eq:ee}
\ee
\subsection*{An Example}
It turns out to be somewhat difficult to obtain a non-trivial contribution to the entanglement entropy coming from (\ref{eq:ee}). The reason lies in the structure of the gauge field $A_i$ (or alternatively, $a_\mu$). $\Delta S_{\mathrm{EE}}$ measures the holonomy of the normal bundle to the bulk entangling surface $\Sigma$. Since the normal bundle involves a time-like direction, this implies that looking for non-zero $\Delta S_{\mathrm{EE}}$ using static metrics on the boundary will give trivial contributions. Thus, although the expression (\ref{eq:entropyformula}) was derived under the assumption that our metric was static, we need to extend this expression to the case where the metric is no longer static. This is  analogous to \cite{Castro} who also extended the entanglement entropy of their topologically massive bulk gravity theory to non-static cases, using \cite{Hubeny} as motivation. They did so even though the proposal of \cite{Hubeny} (and its recent proof \cite{Dong2}) is limited to the case where the bulk action doesn't have higher derivative gravitational terms. We leave the justification for extending (\ref{eq:ee}) to non-static cases as a future project. 

In extending the formula to go beyond the static case, we make the most conservative choice and take as our metric to be a stationary, rotating metric. It turns out, however, that in three dimension such a metric is locally flat. Therefore, it is not a huge stretch for us to use our formula which was derived for the static case.
\subsubsection*{The three-dimensional Kerr solution}
One of the simplest configurations involving a non-static metric that yields a non-trivial contribution to $\Delta S_{\mathrm{EE}}$ is the three-dimensional `Kerr metric' discovered by Deser \emph{et al.\ }in \cite{Deser}. We take as our entangling surface a circle in this geometry. 
The metric is given by 
\be
ds^2_3= -dt^2 +2 b \ dt d\varphi +dr^2+ (r^2-b^2) \ d\varphi^2\, , \label{eq:Kerr}
\ee
where the constant $b$ is proportional to the angular momentum. This metric solves vacuum Einstein's equations. Since in three dimensions Ricci flatness coincides with flatness, this metric is flat. We now discuss a curious feature of this metric as it will be important in interpreting the dual field theory result that we compute in the next section.

As mentioned earlier, this metric is actually locally flat as can be seen by making the coordinate transformation:
\begin{align}
T=t-b \varphi\, .
\end{align}
The metric then becomes
\be
ds^2_3= -dT^2 +dr^2+ r^2 d\varphi^2\, .
\ee
But as $\varphi \rightarrow \varphi+2\pi$, we have $T\rightarrow T -2\pi b$ and so we see that there is a discontinuity in the new time direction. All of this means that even though our `Kerr' metric is flat, it is being sourced by some mass distribution located near $r=0$ with non-zero angular momentum. According to \cite{Deser, Deser2} this metric is sourced by moving point particles (in three dimensions) or moving parallel cosmic strings (in four dimensions). Although here we do not investigate what the sources are, in the next section we shall see that  the field theory `knows' about the presence of the sources.


We now compute the contribution to $\Delta S_{\mathrm{EE}}$ when we take as our entangling surface to be a circle of radius $R$ centred around the origin. Note that the tangent vector to the circle, however, changes from space-like to time-like as the value of $R$ is dialled from greater than $|b|$ to less than $|b|$. Since we want our entangling surface to remain space-like we fix $R>|b|$.

We work with metric expressed in (\ref{eq:Kerr}). Then the normalized vectors which lie normal to the entangling circle are given by
\begin{align}
n^0_\mu &= (n^0_t, n^0_r, n^0_\theta) = \left(1/\sqrt{1 - b^2/r^2}, 0, 0\right)\\
n^1_\mu &= (n^1_t, n^1_r, n^1_\theta) = \left(0, 1, 0\right)\, .
\end{align}
Note that our time-like normal vector becomes imaginary at $r=|b|$. 
It is then straightforward to compute the boundary gauge field $\hat{a}_\mu$,
\begin{align}
\hat{a}_\mu = \left( 0, 0, -\frac{b}{r\sqrt{1 - b^2/r^2}}\right)\, .
\end{align}
Thus, we get the following contribution to the entanglement entropy
\begin{align}
\Delta S_{\mathrm{EE}}= -\frac{\pi\kappa}{4 G}\frac{b}{\sqrt{R^2 - b^2}}\, . \label{eq:kerrentropy}
\end{align}
For comparison with our field theory computation in in the next section we note the small angular momentum or large $R$ limit:
\begin{align}
\Delta S_{\mathrm{EE}}\approx \frac{\pi\kappa b}{4 GR}\, , \label{eq:leadingorder}
\end{align}
where we have dropped the minus sign by replacing $b$ by $-b$. As expected the contribution vanishes in the non-rotating limit. 

\section{Entanglement Entropy}
Now, let's focus on the field theory side. As in the gravity side, we want to compute the contribution to entanglement entropy coming from three dimensional parity violating terms in the quantum field theory.  These are field theories that contain parity-odd terms in the low-energy effective action when coupled to a background gravitational field. The particular term whose effect on the field theory we are considering is the three-dimensional gravitational Chern-Simons term. We get this term on the boundary from the bulk Chern-Pontryagin term,
\begin{align}
  \frac{\kappa}{16\pi G} \int_{B_1} \text{tr} (\Gamma \wedge d\Gamma +\frac{2}{3} \Gamma \wedge \Gamma \wedge \Gamma ) =\frac{\kappa}{32\pi G }\int_{M_1} \text{Tr} (R \wedge R)\, , \label{eq:3dCS}
\end{align}
where $B_1=\partial M_1$. The left-hand-side of this equality plays an essential role in topologically massive theories of gravity \cite{Deser3,Deser4}. Our choice to examine the effect of this term is also justified by \cite{Fischler1}. 

We shall adopt the perturbative approach developed in \cite {Rosenhaus:2014woa, Rosenhaus:2014zza, Hughes:2015ora}  and use the three-dimensional Kerr metric example (\ref{eq:Kerr}) for illustration. Following \cite{Hughes:2015ora}, we interpret our rotating metric as a perturbation of the `flat metric' written in polar coordinates\footnote{In doing the computation we momentarily `forget' that the full rotating metric is locally flat.} as well as analytically continue to Euclidean signature. The latter, of course, also entails making the angular momentum parameter $b$ to be imaginary.

As a starting point we note that $\Delta S_{\mathrm{EE}}=0$ for a circular entangling surface on a flat spacetime. We then perturb the metric which leads to the following expression for $\Delta S_{\mathrm{EE}}$,
\be \label{variation}
\Delta S_{\text{EE}}=\frac{1}{2} \int_{\mathbb{R}^3} d^3 x \delta g^{\mu \nu} \langle T_{\mu \nu} (\bold {x}) \widehat{H} {\rangle} _\mathrm{conn}\,  ,
\ee
where $ \langle ... \rangle_\mathrm{conn} $ is the connected two point function. $T_{\mu \nu}$ is the unperturbed energy momentum tensor of the QFT and $\widehat{H}$ is the modular Hamiltonian. $\delta g^{\mu \nu}$  is the perturbation around the background geometry. For a two-dimensional ball of radius $R$ centred at the origin, the modular hamiltonian is given by\cite{Casini:2011kv}
\be
\widehat {H}= 2 \pi \int_0^R dr' \int_0^{2\pi} d\theta' r' \left(  \frac{ R^2-r'^2 }{2R} \right) T_{00} (\tau', r',\theta ') +\text{constant}\, ,
\ee
where the constant in the above equation is there to ensure that the density matrix is normalized to unity. It does not play a role in the connected part of the correlation function and so we drop it below.

In \cite{Closset:2012vp}, the parity odd contribution to the energy-momentum two-point function coming from the term on the left-hand-side of (\ref{eq:3dCS}) was computed to be
\be \label{eq:TT}
\langle T_{\mu\nu} (x) T_{\lambda\rho}(x')\rangle = - \frac{i \kappa'} {16 \pi}\ \epsilon_{(\mu (\lambda \sigma} \left( \nabla_{\nu) \rho)} - g_{\nu)\rho)} \nabla^2 \right) \nabla^\sigma \frac{\delta^3(x-x')}{\sqrt{g}}\, ,
\ee
where $\kappa'$ is a dimensionless constant and it is related to the bulk quantity by
\begin{align}
\kappa' = \frac{\kappa}{G}\, .
\end{align}
In (\ref{eq:TT}) we have `covariantized' the expression since we are working in polar coordinates in which the modulus of the determinant of metric tensor is not unity. We note that
\be
\delta g^{\tau \theta} = \frac{b}{r^2} \;\mathrm{and}\; \delta g^{\theta\theta} = \frac{b^2}{r^4}\, .
\ee
We compare our computation with the leading order result in the holographic computation (\ref{eq:leadingorder}) and so we only consider the $\delta g^{\tau\theta}$ perturbation.

Plugging in the values and doing the $\tau$ and the $\theta'$ integrals we get the following formal expression for $\Delta S_{\mathrm{EE}}$:
\begin{align}
\Delta S_{\mathrm{EE}}= \frac{ \kappa' i b}{8 } \int_\varepsilon^{\infty} dr \int_0^{2\pi} d\theta \int_0^R dr' r' \left(\frac{R^2 - r'^2}{2R}\right)  \left(\frac{1}{r}\partial^2_r+\partial^3_r+ \frac{1}{r^2} \partial^2_{\theta} \partial_r \right) \frac{\delta(r-r')}{r}\, .
\end{align}
In the above expression we have introduced a lower cut-off $\varepsilon$ to the $r$ integral since the integral has a singularity coming from the origin. After performing the delta function integrals carefully we get: 
\be
\Delta S_{\mathrm{EE}} =\frac{\pi i \kappa'}{4} \left(\frac{b}{R} - \frac{b}{2R}\ln{R/\varepsilon}\right)\, . \label{eq:fieldtheoryresult}
\ee
In the above expression we get a factor of $i$ due to the fact that the computation was done in Euclidean space in which the angular momentum $b$ has to be taken to be imaginary for the metric to remain real. We see that (\ref{eq:fieldtheoryresult}) agrees with the holographic computation (\ref{eq:leadingorder}) in the first term but that in the field theory we get an extra divergence from the origin. 

In the previous section, we saw that the background metric had a curious discontinuity in the time direction when written in `flat' coordinates. This accounts for the fact that there was a rotating matter source near the origin of the spatial sections. In our computations we did not specify this source. Since on the field theory side we see a short-distance cutoff from the same region of spacetime, we interpret this divergence as encoding our ignorance of the matter source that creates this rotating spacetime. In other words, the field theory `knows' about the rotating source and assigns some cut-off dependent entropy to the source. In a scenario in which one specifies the sources on both sides, we believe that one should be able to do the matching of the entanglement entropies.

\section{Conclusion}
In the spirit of understanding the intimate relationship between quantum entanglement and gravity, we have devoted this note to computing the entanglement entropy of Jackiw and Pi's modification of general relativity. As described in the introduction, the modification is affected through the addition of a Chern-Pontryagin density. Unless its coefficient is promoted to a dynamical spacetime scalar, this term is topological and does not contribute to the gravitational dynamics. However, as we have demonstrated, it does contribute to the entanglement entropy through a term proportional to the curvature flux through the normal bundle to the Ryu-Takayanagi entangling surface. In this sense, our computation may be considered further evidence in support of Dong's proposal for gravitational entropy in higher derivative gravity. We have exemplified the formal argument in the case of a stationary, rotating background and, following the perturbative approach of \cite {Rosenhaus:2014woa, Rosenhaus:2014zza, Hughes:2015ora}, matched this with the entanglement entropy of a (generic) three-dimensional field theory with parity violating terms that contribute to the universal topological entropy. The match is not perfect. In addition to the anticipated part of $\Delta S_{EE}$, we find a second term that diverges like the log of the cut-off scale. We speculate that this term codes the matter source at the origin of the bulk spacetime in some way. This would be an interesting future avenue to explore. 

\section{Acknowledgements}
The authors would like to thank \'Alvaro V\'eliz-Osorio, Pawe\l{} Caputa, David Kubiz\v{n}\'ak, Michal Heller and Aitor Lewkowycz for useful discussions and comments on the manuscript. We would also like to thank Arpan Bhattacharyya for collaboration at the earlier phase of this work and Onkar Parrikar for correspondences. TA and SSH would also like thank the hospitality of the organizers of the Simons Summer Workshop 2015 where this project was initially conceived. JM gratefully acknowledges support by NSF grant PHY-1606531 at the Institute for Advanced Study and NRF grant GUN 87667 at the University of Cape Town. SSH is supported by the Claude Leon Foundation. TA's research is funded by the Perimeter Institute for Theoretical Physics. Research at Perimeter Institute is supported by the Government of Canada through Industry Canada and by the Province of Ontario through the Ministry of Economic Development and Innovation.

\bibliography{references}

\bibliographystyle{utphysmodb}

\end{document}